\documentclass{elsart}
\usepackage[dvips]{graphicx}

\begin{document}

\begin{frontmatter}
\title{Alkali Halide Surfaces Near Melting: Wetting and Nanofriction 
       Properties}
\author[sissa,democritos]{D. Ceresoli},
\author[sissa,democritos]{T. Zykova-Timan\thanksref{now}},
  \thanks[now]{Present address: Computational Science Department of
  Chemistry and Applied Biosciences, ETH Zurich USI-Campus, Via Giuseppe
  Buffi 13, LUI CH-6900 Lugano, Switzerland}
\author[sissa,democritos]{U. Tartaglino} and
\author[sissa,democritos,ictp]{E. Tosatti\corauthref{corresponding}}
   \corauth[corresponding]{Corresponding author.
   \emph{Phone:} +39--040--3787--438, \emph{Fax:} +39--040--3787--528,}
   \ead{tosatti@sissa.it}
\address[sissa]{Scuola Internazionale Superiore di Studi Avanzati (SISSA),
  via Beirut 2-4, 34014 Trieste, Italy}
\address[democritos]{DEMOCRITOS National Simulation Center, via Beirut 2-4,
   34014 Trieste, Italy}
\address[ictp]{International Center for Theoretical Physics (ICTP),
   Strada Costiera 11, 34014 Trieste, Italy}

\begin{abstract}
Alkali halide (100) crystal surfaces are poorly wetted by their own melt
at the triple point. We carried out simulations for NaCl(100) within
the well tested BMHFT model potential. Calculations of the solid-vapor,
solid-liquid and liquid-vapor free energies showed that solid NaCl(100)
is a non-melting surface, and explain its bad wetting in detail.
 The extreme stability of NaCl(100) is ideal for
a study of the nanofriction in the high temperature regime, close to and
even above the bulk melting temperature ($T_\mathrm{M}$).  Our simulations
reveal in this regime two distinct and opposite phenomena for plowing and
for grazing friction.  We found a frictional drop close to $T_\mathrm{M}$
for deep ploughing and wear, but on the contrary a frictional rise
for grazing, wearless sliding.  For both phenomena we obtain a fresh
microscopic understanding, relating the former to ``skating'' through
a local liquid cloud, the latter to softening of the free substrate
surface. It is argued that both phenomena, to be pursued experimentally,
should be much more general than the specific NaCl surface case. Most
metals in particular possessing one or more close packed non-melting
surface, such as Pb, Al or Au(111), should behave quite similarly.
\end{abstract}

\end{frontmatter}

\section{Introduction}
The high temperature surface properties of alkali halide crystals,
particularly of NaCl, are very unusual as these solids are incompletely
wetted by \emph{their own} melt.~\cite{grange75} We investigated
theoretically these surfaces and their uncommon behavior, with a multiple
goal. The first goal was to uncover the physical reasons for the poor
wetting. The second was to exploit the availability of the standard
empirical potential parametrized long ago by Tosi and Fumi (BMHFT)
for a first quantitative characterization of all the interfaces --
and particularly their structure and their temperature dependent
interface free energies -- in a well defined solid-liquid-vapor
system. Achieving this kind of goal is of particular interest in the
general context of high temperature capillarity, a field where such a
detailed understanding is usually unavailable. A third goal was to verify
whether the solid NaCl(100) surface exhibits surface non-melting, as the
partial self-wetting would imply on thermodynamic grounds. A fourth goal
was to study the NaCl liquid surface, extracting from its temperature
dependent surface tension the surface entropy, and comparing it. e.g.,
with the solid surface entropy at the melting point. Again, it seems
that this comparison is generally unavailable. A fifth and more practical
type of goal was finally to put to use the understanding obtained for the
non-melting solid surface, in the field of nanofriction. In particular,
our aim is to provide a first theory and simulation approach to both
\emph{ploughing} and \emph{grazing} friction of hard nanotips on a solid
surface near melting.

In the following we summarize the results of work that appeared in
Refs.~\cite{droplet04,PRL05,JCP05,NatMat07}.

\section{Partial wetting of NaCl by its own melt}
\begin{figure}\begin{center}
  \includegraphics[width=0.45\textwidth]{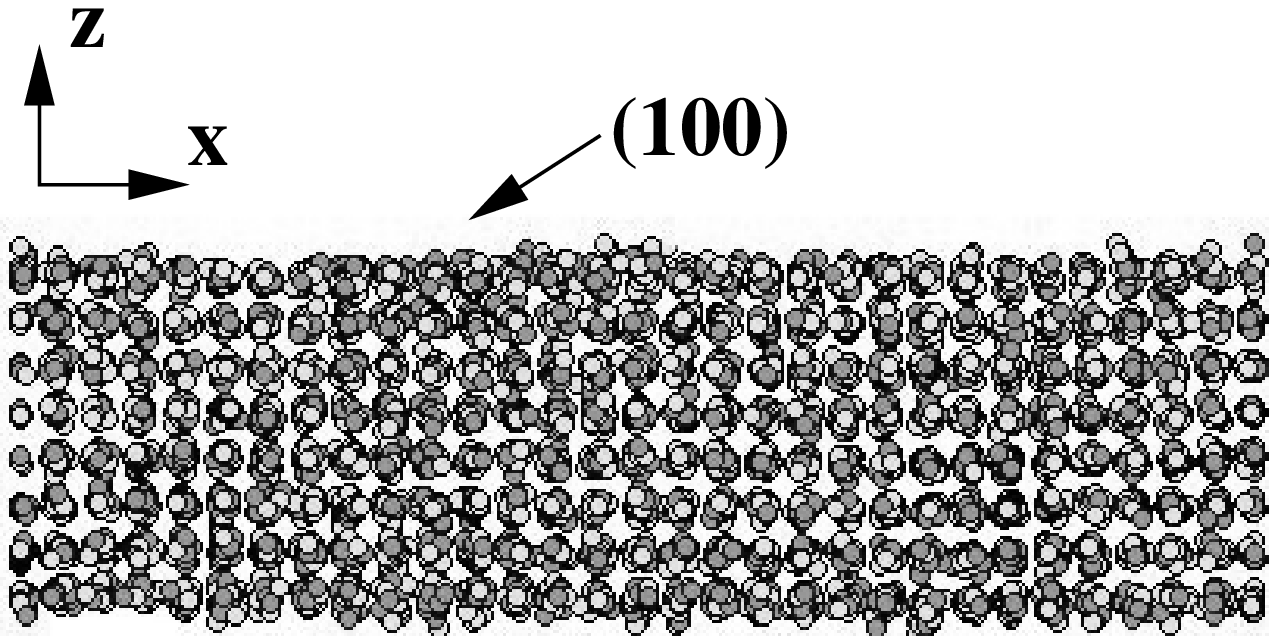}
  \caption{Snapshot of a simulation of a NaCl(100) slab, at 1120~K~$\simeq
  T_\mathrm{M}+50$~K (the calculated bulk melting temperature is
  $T_\mathrm{M}=$~1066~K, in good agreement with experimental data).
  The (100) surface is normal to the $z$ direction, indicated in
  the figure.  In our simulations at 1120~K, NaCl(100) remains
  crystalline in a metastable state for at least 200~ps, confirming
  its non-melting nature. Upon increasing temperature, crystalline
  NaCl(100) persists up to a \emph{surface spinodal temperature}
  $T_\mathrm{ss}\simeq$~1215~K~$\simeq T_\mathrm{M}+150$~K.}
  \label{fig1}
\end{center}\end{figure}

\begin{figure}\begin{center}
  \includegraphics[width=0.45\textwidth]{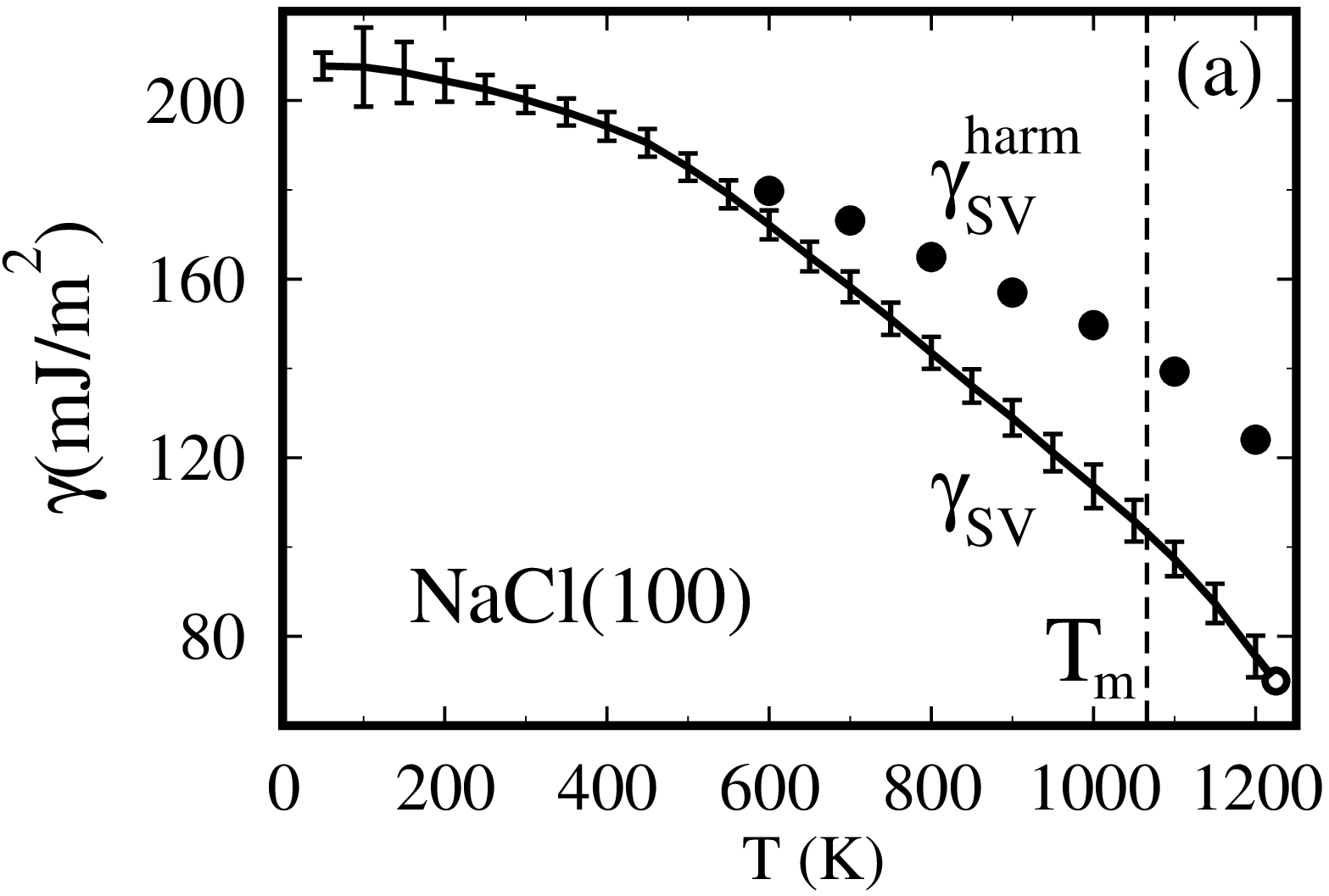}\\
  \includegraphics[width=0.45\textwidth]{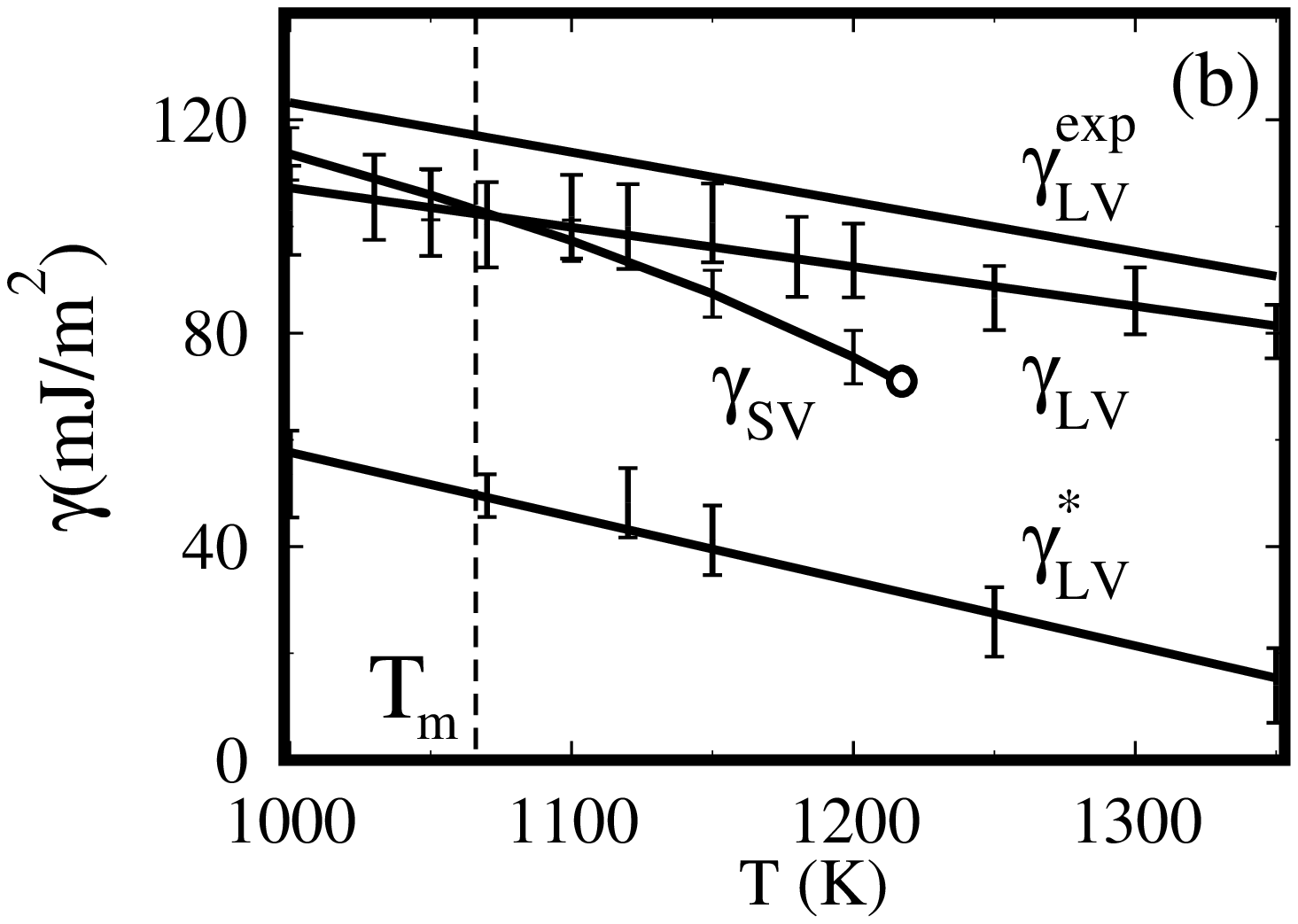}
  \caption{(a) The solid surface free energy
  $\gamma_\mathrm{SV}$ from thermodynamic integration (solid line)
  and from the effective harmonic approximation (dots)~\cite{JCP05}.
  (b) Liquid NaCl surface tension $\gamma_\mathrm{LV}$,
  compared to the experimental value $\gamma_\mathrm{LV}^\mathrm{exp}$.
  The solid surface free energy $\gamma_\mathrm{SV}$ is reported in
  the same plot. $\gamma^\star_\mathrm{LV}$ is the surface tension
  re-calculated by artificially removing the correlations between atoms
  at the outer surface.~\cite{PRL05,JCP05} Once surface molecular order
  is removed in this way, the surface entropy rises and the surface
  tension drops. Solid NaCl(100) would be completely wet by this
  artificial liquid.}
  \label{fig2}
\end{center}\end{figure}

\begin{figure}\begin{center}
  \includegraphics[width=0.45\textwidth]{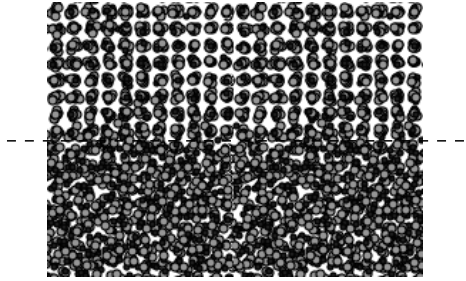}
  \caption{Simulated NaCl(100) solid--liquid interface at the melting
  point.  The interface is quite sharp. Its free energy is calculated
  to be $\gamma_\mathrm{SL}=$~36$\pm$5~mJ/m$^2$.}
  \label{fig3}
\end{center}\end{figure}

\begin{figure}\begin{center}
  \includegraphics[width=0.45\textwidth]{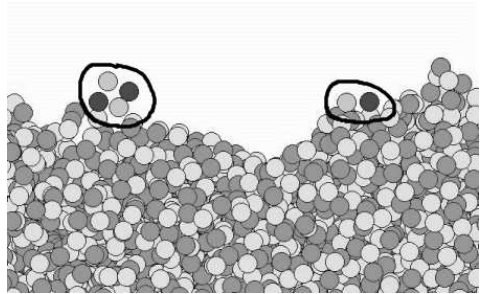}
  \caption{Snapshot of a simulation of a liquid NaCl surface
  (liquid--vapor interface), at 1250~K well above the melting point and
  even above the surface spinodal temperature $T_\mathrm{ss}$.  Note the
  very pronounced fluctuations in the instantaneous surface profile.
  This picture, suggestive of a low surface tension, high entropy surface,
  is in apparent contradiction with the massive non-wetting of solid
  NaCl(100).  Indeed, we found in simulation~\cite{PRL05,JCP05}, an
  \emph{incipient molecular surface ordering} which lowers the surface
  entropy by a factor $\simeq$~2.5.  The origin of this molecular
  surface order is charge order, which, already important in bulk,
  plays an enhanced role at the molecular liquid surface. Analyzing
  the correlations between Na$^+$ and Cl$^-$ ions at the surface, one
  finds that the outermost layer is rich in NaCl monomers and dimers,
  as highlighted in the figure.}
  \label{fig4}
\end{center}\end{figure}

Our classical molecular dynamics simulations with the Tosi-Fumi
potentials confirm that in this model crystalline NaCl(100) is only partly
wetted by a (nano)~droplet of molten NaCl.~\cite{droplet04} Moreover,
as expected, solid NaCl(100) is confirmed to be a non-melting surface,
stable without any precursor signals of melting up to the bulk melting
point. In a metastable state, and in the absence of defects, it can
even be overheated by as much as 100K above the melting temperature
(Fig.~\ref{fig1}). Extracting from the simulations the properties of
the three interfaces -- solid-vapor, liquid-vapor, and solid-liquid
-- we calculated and analyzed their free energies (Fig.~\ref{fig2})
and found that the surface non-melting of NaCl(100) and the resulting
partial self-wetting stem from three separate reasons:
\begin{itemize}
\item[(i)] Solid NaCl(100) is an exceptionally stable solid surface
(Fig.~\ref{fig1}), with a strongly decreasing surface free energy at high
temperature (Fig.~\ref{fig2}). Stability in this regime is enhanced by
extremely large anharmonicities. At the melting point $T_\mathrm{M}$,
we found $\gamma_\mathrm{SV}=$~103~mJ/m$^2$.
\item[(ii)] The solid-liquid interface is spatially
sharp (Fig.~\ref{fig3}), and is energetically very expensive
($\gamma_\mathrm{SL}\simeq$~36~mJ/m$^2$). This can be attributed to large
structural differences between solid and liquid, in particular a 27\%
density difference between the two.
\item[(iii)] The liquid-vapor interface free energy (surface tension)
is relatively high ($\gamma_\mathrm{LV}\simeq$~102~mJ/m$^2$), actually
almost identical to the solid-vapor free energy at the melting point. The
reason for this high surface tension is traced to a surprising {\em
deficit} of liquid surface entropy, which we calculate to be a factor
$\simeq$2.5 {\em lower} than that of the solid surface. Low entropy
implies some kind of short range order at the molten salt surface. We
highlighted and characterized in particular an incipient pairwise
molecular charge order in the outermost regions of the molten salt surface
(Fig.~\ref{fig4}).~\cite{PRL05,JCP05}
\end{itemize}

\section{High temperature nanofriction}
The unusual stability of the alkali halide surface against melting suggests
adopting it as a natural toy system for the study of high temperature
AFM tip-surface nanofriction studies. A non-melting surface will not
automatically liquefy under a tip, (as would instead be the case for a
regular, melting surface~\cite{tomagnini93}) even at temperatures very
close to bulk melting. For this reason, NaCl(100) and other alkali halide
surfaces should provide good testing grounds for a first study of high
temperature nanofriction, a potentially important area which is still
largely unexplored.  In the absence of any experimental data on high
temperature nanofriction, we conducted exploratory molecular dynamics
sliding friction simulations of hard (diamond) tips on NaCl(100), both
in the heavy ploughing, wear-dominated regime, and in the light grazing,
wearless regime.

\begin{figure}\begin{center}
  \includegraphics[width=0.45\textwidth]{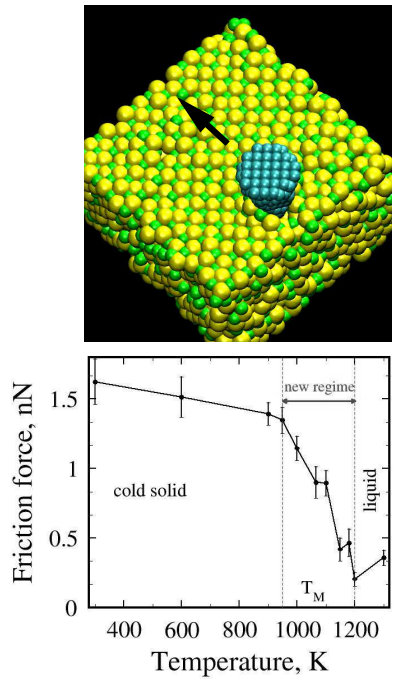}
  \caption{Top panel: initial stage picture of ploughing friction
  of a hard, sharp tip on NaCl(100), 100 degrees below the melting
  point. The tip is modeled as a rigid diamond apex, 26~\AA\ long.
  The tip indentation depth is $\simeq$~6~\AA. Note the fast-healing
  furrow behind the tip.  Bottom panel: averaged ploughing frictional
  force $\langle F_x \rangle$ as a function of temperature. Note the
  friction drop at around $T = T_\mathrm{M}-$150~K, which is due to
  ``skating'' of the tip.}
  \label{fig5}
\end{center}\end{figure}

The simulated ploughing friction of a sharp penetrating tip is very large,
as expected for scratching a hard solid (Fig.~\ref{fig5}). However,
it shows as a function of temperature a strong drop near the melting
point. This high temperature friction decrease can be assimilated to
\emph{skating} of the tip over the hot solid. As in ice skating, the
tip is surrounded by a local liquid halo, moving along with the tip as
it ploughs on.~\cite{NatMat07}

\begin{figure}\begin{center}
  \includegraphics[width=0.45\textwidth]{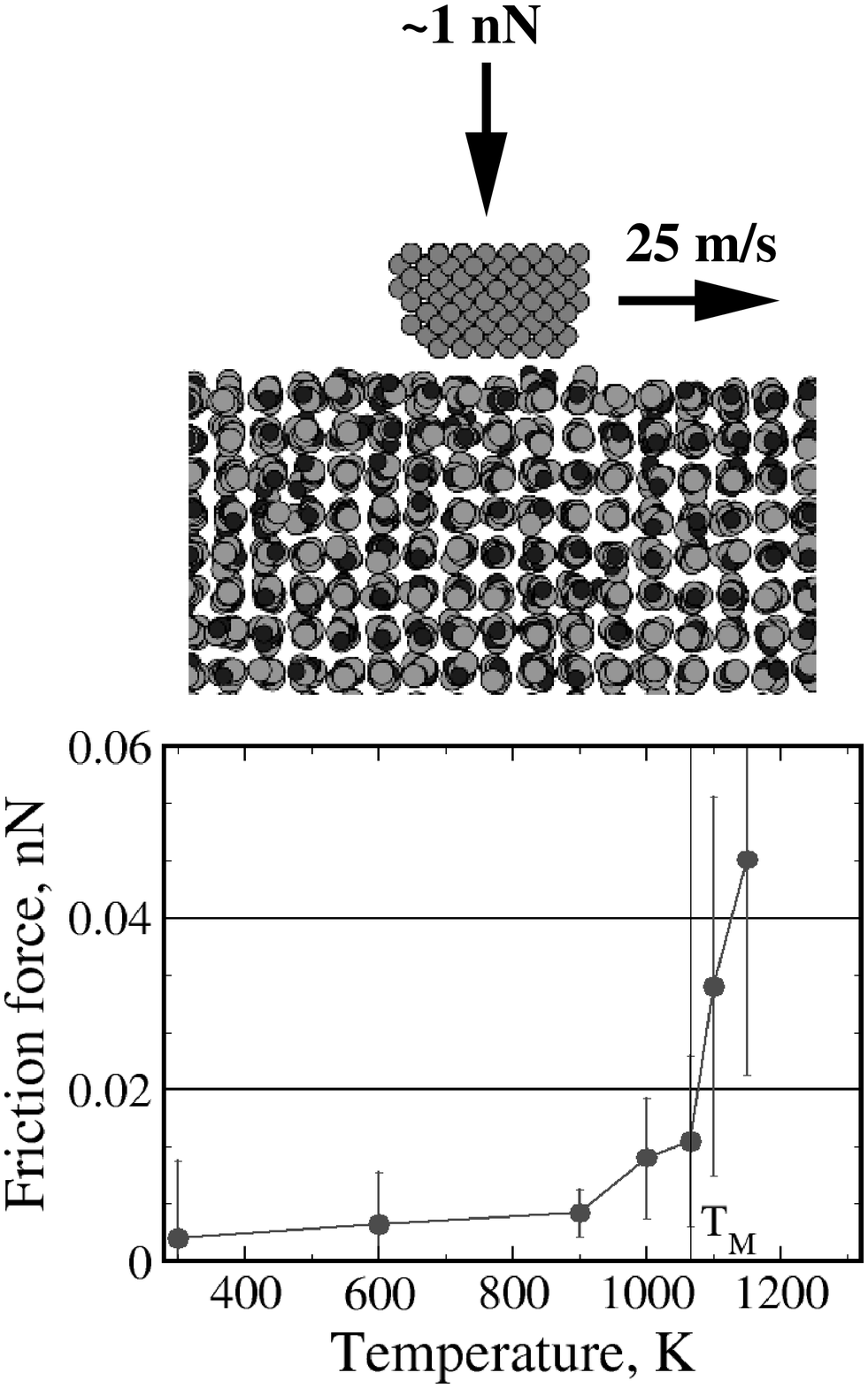}
  \caption{Top panel: snapshot a grazing friction simulation. The
  tip is modeled as rigid hard diamond tip, exposing a flat facet of
  about 132~\AA$^2$. The tip contacts the substrate by mere adhesion
  forces, which we estimate to be $\simeq$~1~nN.  Bottom panel:
  average frictional force of the grazing tip on NaCl(100).  Note the
  initially very low friction (due to incommensurability between the two
  lattices), however sharply increasing at $T_\mathrm{M}$ and above. Note
  the contrast with the ploughing friction drop of Fig.~\ref{fig5}}
  \label{fig6}
\end{center}\end{figure}

At the opposite extreme, we found that simulated grazing friction
of a hard flat tip over NaCl(100) behaves just the other way around
(Fig.~\ref{fig6}). Unlike ploughing, low temperature grazing friction is
initially very small, on account of the incommensurability of the two
respective lattice planes.  Upon heating however the grazing friction
surges just near the melting point. This rise is seen as the analog of
the celebrated \emph{peak effect} in the critical current close to the
upper critical field in the mixed state of type II superconductors. The
dramatic softening of the flux lattice just before its disappearance
causes a sudden increase of sliding friction of fluxons against the
pinning impurities.~\cite{chaikin96,granato2000} In our case the softening
of the NaCl surface lattice, still solid at the melting temperature,
reflects in an increase of sliding friction just before spontaneous
breakdown of the solid surface above the melting point.~\cite{NatMat07}

\section{Conclusions}
This contrasting behavior of friction close to the melting point --
skating when scratching, braking when grazing -- should be more general
than the specific instance of NaCl(100) where we uncovered them. Metals
in particular usually possess at least one close packed non-melting
surface~\cite{report05} such as the (111) or the (110) surfaces of
\emph{fcc} or \emph{bcc} metals respectively, which is non-melting and
should behave similarly to NaCl(100). We moreover can expect possible
modifications of the high temperature nanofriction behavior by adhesive
forces between tip and surface, as well as by application of a bias
between (conducting) tips and surfaces.

\section*{Acknowledgments}
This work was sponsored by MIUR and COFIN 2006022847, as well as by
INFM (Iniziativa trasversale calcolo parallelo).



\clearpage

\clearpage

\clearpage

\clearpage

\begin{thebibliography}{99}
\bibitem{grange75}  
  G. Grange, B. Mutaftschiev,
  Surf. Sci. \textbf{47} (1975) 723.

\bibitem{droplet04}
  T. Zykova-Timan, U. Tartaglino, D. Ceresoli, W. Sekkal-Zaoui,
  E. Tosatti, Surf. Sci. \textbf{566} (2004) 794.

\bibitem{PRL05}
  T. Zykova-Timan, D. Ceresoli, U. Tartaglino, E. Tosatti, 
  Phys. Rev.Lett. \textbf{94} (2005) 176105.

\bibitem{JCP05}
  T. Zykova-Timan, D. Ceresoli, U. Tartaglino, E. Tosatti,
  J. Chem. Phys. \textbf{123} (2005) 164701.

\bibitem{NatMat07}
  T. Zykova-Timan, D. Ceresoli, E. Tosatti,
  Nature Materials \textbf{6}, 230 (2007).

\bibitem{tomagnini93}
  O. Tomagnini, F. Ercolessi, E. Tosatti,
  Surf. Sci. \textbf{287/288} (1993) 1041.

\bibitem{chaikin96}
  C. Tang, X. Ling, S. Bhattacharya, P. Chaikin,
  Europhys. Lett \textbf{35} (1996) 597.

\bibitem{granato2000}
  E. Granato, T. Ala-Nissia, S. Ying,
  Phys. Rev. B \textbf{62} (2000) 11834, and references therein.

\bibitem{report05}
  U. Tartaglino, T. Zykova-Timan, F. Ercolessi, E. Tosatti,
  Phys. Repts. \textbf{411} (2005) 291.

\end{thebibliography}
\end{document}